\documentstyle[epsfig,aps]{revtex}
\input epsf
  
\newcommand{\be}{\begin{equation}}
\newcommand{\ee}{\end{equation}}   

\newcommand{\bea}{\begin{eqnarray}}
\newcommand{\eea}{\end{eqnarray}}

\begin{document}
\title{\Large Color Screening and the Suppression of the Charmonium State
Yield in Nuclear Reactions}

\author{L. Gerland${}^a$, L. Frankfurt${}^{a}$, M. Strikman${}^b$,
H. St\"ocker${}^c$\\
${}^a$ School of Physics and Astronomy, Tel Aviv University,\\
69978 Ramat Aviv, Tel Aviv, Israel
\\
${}^b$ Department of Physics, Pennsylvania State University\\
University Park, PA 16802, USA
\\
${}^c$ Institut f\"ur Theoretische Physik der 
J.W.Goethe-Universit\"at\\
Robert-Mayer-Str. 8-10, D-60054 Frankfurt, Germany}

\date{\today}

\maketitle
\begin{abstract}

We discuss the new data for the production of the $\psi'$ meson in pA
collisions at 450 GeV at CERN-SPS (of the NA50-collaboration)~\cite{na50}.  
We extract from the CERN data $\sigma(\psi' N)\approx 8$~mb under the
assumption that the $\psi'$ is produced as a result of the space-time
evolution of a point-like $c\bar c$ pair which expands with time to the full
size of the charmonium state.  In the analysis we assume the existence of a
relationship between the distribution of color in a hadron and the cross
section of its interaction with a nucleon.  However, our result is rather
sensitive to the pattern of the expansion of the wave packet and 
significantly larger values of $\sigma(\psi' N)$ are not ruled out by the
data. We show that recent CERN data confirm the suggestion of
ref.~\cite{ger} that color fluctuations of the strengths in
charmonium-nucleon interaction are the major source of suppression of the
$J/\psi$ yield as observed at CERN in both pA and AA collisions.

\end{abstract}

\section{Introduction}

The aim of the present paper is to show that the theoretical description
of the production and annihilation yield of particles with hidden charm,
which accounts for color fluctuations in charmonium-nucleon interactions,
agrees with new data of the NA50-collaboration~\cite{na50} who observed
quite different $J/\psi$ and $\psi'$-nucleon cross sections.  Analogous
results have been found at negative Feynman x ($x_F$) in proton-nucleus
collisions at 800 GeV at Fermilab by the E866 collaboration~\cite{leitch}:
Both experiments found that the charmonium-nucleon cross section is
smaller in the target fragmentation region than at midrapidity. This is in
good agreement with models that assume that the $\psi'$ is produced due to
a space-time evolution of colorless, point-like $c\bar c$ pairs, which
expands with time to its full size and that there exists a relationship
between the distribution of color in a hadron and the cross section of its
interaction with a nucleon.  The agreement of this scenario with the
negative $x_F$ Fermilab data~\cite{leitch} with was first demonstrated in
ref.~\cite{ger}. 

The major new effect is that QCD predicts different interaction cross
sections for the interaction of different charmonium states ($J/\psi$,
$\psi'$ and $\chi$) with usual hadrons due to their different spatial size
if the cross section is proportional to the distribution of color in the
projectile. This relationship is proved in pQCD (perturbative
QCD)~\cite{sigma} as equivalent of the QCD evolution equations and it is a
plausible suggestion in the nonperturbative QCD regime, which assumes smooth
matching with pQCD predictions. If this is so, the $\psi'$-nucleon cross
section is expected to be on the order of the interaction cross section of
nucleons with mesons built of light quarks, e.g. the $\phi$ or the $K$. 

It is a well known fact that spatially larger hadrons have larger
interaction cross sections. For example, the inelastic $J/\psi$-nucleon
cross section was found to be around 3.5 mb at SLAC~\cite{slac} while the
inelastic $\pi$-nucleon cross section at these energies is approximately
20 mb. From charmonium models, e.g. in ref.~\cite{eich,quigg} it is known
that the different charmonium states ($J/\psi$, $\chi$ and $\psi'$) have
different spatial sizes. The average relative
distances $\sqrt{\langle r^2\rangle}$
with $\langle r^2\rangle=\int \Psi^2(r)\cdot r^2 {\rm d}^3r$ of the $Q\bar
Q$-pairs are $\sqrt{\langle r^2\rangle}_{J/\psi}=0.38$~fm, $\sqrt{\langle
r^2\rangle}_{\chi}=0.57$~fm and $\sqrt{\langle r^2\rangle}_{\psi'}=0.76$~fm
for the wave functions $\Psi(r)$ of ref.~\cite{eich}.  

The relation between the spatial size of a color dipole and the interaction
cross section was found in pQCD~\cite{sigma} for a spatially small dipole as
another form of the QCD evolution equations. It will be shown in
sect.~\ref{hardxsect} that the charmonium-nucleon cross section predicted
from pQCD is significantly smaller than the measured ones. We concluded in
ref.~\cite{ger} that these cross sections are dominated by non-perturbative
contributions. A plausible parametrization within the constituent quark
model for this soft contribution is reviewed in section~\ref{appendix}.

In ref.~\cite{ger} it was shown that the production of $J/\psi$'s in pA
collisions can be understood if one takes into account the production and
the subsequent decays of higher mass charmonium state resonances
($\chi$,$\psi'$) into $J/\psi$'s. Those higher resonances have larger
cross sections for the scattering off nucleons due to their larger spatial
size. This leads to a significant increase of the effective
absorption of $J/\psi$'s as compared to the propagation
of pure $J/\psi$-states. In ref.~\cite{spieles} it was shown also that the
production of $J/\psi$'s in AA collisions is additionally suppressed by the
final state interaction of charmonium states with newly produced hadrons
like $\pi$'s, $\rho$'s and so on. These interactions of charmonium
states with light hadrons are predominantly soft in the SPS-regime,
because they take place at lower energies than the collisions with the
initial nucleons.
 
\section{Model description}

\subsection{Semiclassical Glauber approximation \label{model}}

The suppression factor $S$ for the charmonium production in 
minimum bias $pA$ collisions can be evaluated within the semiclassical
approximation (cf.~\cite{yennie}) as
\begin{equation}
S_A=\frac{\sigma(pA\rightarrow X)}{A \cdot \sigma(pN\rightarrow X)} =
{1\over A} \int {\rm d}^2B\,{\rm d}z\, \rho(B,z)\cdot \exp \left(-
\int_z^{\infty}\sigma(XN)\rho(B,z'){\rm d}z'\right)\;.
\label{glaub}
\end{equation}

$\rho(B,z)$ is the local nuclear ground state density (we used the
standard parametrization from~\cite{devries}). $\sigma(XN)$ is the
interaction cross section of the charmonium state $X$ with a nucleon.  We
want to draw attention to the fact that this cross section changes with
time due to the space-time evolution of color fluctuations. Therefore, it
is necessary to keep $\sigma$ under the integral.  In principle one should
deal with the expansion effects on the amplitude level. However,
numerically the difference is small on the scale of the uncertainties of
the modeling of the effect.

We make the standard assumption that the production of $c\bar c$ pairs
is a hard process and that the QCD factorization theorem is applicable,
similar to the Drell-Yan process. In practice this means for Drell Yan pair
production that $S=1$, because $\sigma(pA\rightarrow l^+l^-)= A\cdot
\sigma(pp\rightarrow l^+l^-)$.  Deviations from $S=1$ for the Drell-Yan
process can result from nuclear effects on the parton distribution only,
which are neglected, because in the kinematics of SPS they are a small
correction. However, in contrast to the dilepton pair produced in the
Drell-Yan process, the charmonium states have strong final state
interactions. They can be split into open charmed hadrons by colliding with
other hadrons. Therefore, $0<S<1$ for the charmonium states. 

The $S=1$ assumption on the level of the charm quark production leads
to a restriction on the range of applicability of the discussed model
as the gluon distribution is modified at small x due to the shadowing
and antishadowing effects. At the very least the requirement is
that for very high  energies
\begin{equation}
E_{J/\psi}\le {m^2_{J/\psi} \over 2x_0\cdot x_P m_N},
\end{equation}
where $x_0\sim 0.03$ is in the region where gluon shadowing sets in.
$x_P$ is the Bjorken x of the projectile. 
For $y_{c.m.}=0$ the restriction 
corresponds to  $x_0=m_{J/\psi}/2\sqrt{s}$, because $x_P=x_T$ with $x_T$ 
is the Bjorken x of the target.
In addition, as soon as $x_A$ of the nuclear gluon becomes small enough
the $c\bar c$ pair is produced at a distance $\sim 1/(2 x_Am_N)$ from the
interaction point and one has to take into account the evolution of the pair 
before it reaches the interaction point.
Hence the picture of the formation of $c\bar c$ states in the proton 
fragmentation region is qualitatively different from that in the nuclear 
fragmentation region.

The suppression factor $S$ of $J/\psi$'s produced in the nuclear medium is
calculated as:
\begin{equation}
S=0.6\cdot ( 0.92\cdot S^{J/\psi}+0.08\cdot S^{\Psi'})+0.4\cdot
S^{\chi}\; .
\label{mix}
\end{equation}

$S^X$ are the respective suppression factors of the different pure
charmonium states $X$ in nuclear matter.  Eq.~(\ref{mix}) accounts for the
decay of higher resonances after they left the target nucleus into
$J/\psi$'s. The fractions of $J/\Psi's$ that are produced in the decays of
higher resonances in eq.~(\ref{mix}) are taken from ref.~\cite{kharzeev}.
However, in ref.~\cite{kharzeev} it is assumed that the different charmonium
states interact with nucleons with the same cross section, which is in
disagreement with the data from the refs.~\cite{na50,leitch}.

In line with the above discussion we want to stress that
Eqs.~(\ref{glaub})~and~(\ref{mix}) are applicable at CERN energies for
central and negative rapidities, but have to be modified if applied already
at $y_{c.m.}\sim 0$ at RHIC or higher energies, because at higher energies
charmonium states can be produced outside of the nucleus and the $c\bar c$
pairs propagate through the whole nucleus without forming a hadron.

Data are often presented in the form:
\begin{equation} 
\sigma_{pA}=\sigma_{pp}\cdot A^\alpha 
\label{alpha}
\end{equation}
The relation between $S$ and $\alpha$ is
\begin{equation}
S=A^{\alpha-1}\;.
\end{equation}

\subsection{Color Fluctuations in Charmonium Rescattering
\label{model2}}

The first evaluations of $\sigma_{tot}(J/\psi-N)$ have been obtained by
applying the Vector Dominance Model (VDM) to $J/\psi$ photoproduction data.
This leads to $\sigma_{J/\psi N} \sim 1$ mb for $E_{inc} \sim 20$ GeV.
However, the application of the VDM leads to a paradox~\cite{FS85} -- one
obtains $\sigma_{tot}(\Psi'-N) \approx 0.7\cdot \sigma_{tot}(J/\psi-N)$,
although, on the other hand, $r_{\Psi'} \approx 2 r_{J/\psi}$ in charmonium
models like in the refs.~\cite{quigg,eich}. This clearly indicates that the
charmonium states produced in photoproduction are in a smaller than
average configuration. Therefore, the VDM significantly underestimates
$\sigma_{tot}(J/\psi-N)$ and $\sigma_{tot}(\psi'-N)$~\cite{FS85}.

Note that the Generalized VDM, which assumes that the dominant process is
photoproduction of spatially small $c\bar c$ pairs and accounts for the 
space time evolution of this pair, predicts significantly larger
$\sigma_{tot}(\psi'-N)$~\cite{fs91}.

Indeed, the $A$-dependence of the $J/\psi$ production studied at SLAC at
$E_{inc} \sim 20$ GeV exhibits a significant absorption effect~\cite{slac}
corresponding to $\sigma_{abs}(J/\psi-N)= 3.5 \pm 0.8$ mb. It was
demonstrated in \cite{farrar} that, in the kinematic region at SLAC, the
effects due to the space-time evolution of the $J/\psi$ are still small for
the formation of $J/\psi$'s and lead only to a small increase of the value
of $\sigma_{abs}(J/\psi-N)$.  So, in contrast to the findings at higher
energies, at intermediate energies this process measures the {\it genuine}
$J/\psi-N$ interaction cross section at energies of $\sim $ 15-20 GeV
\cite{farrar}. However, the dynamical effect of the production of charmonium
states in squeezed configurations is still there. We account for this
effect as due to the propagation of the $J/\psi$ and $\psi'$ system. 
Experimental evidence for this expansion was found e.g.\ in the
photoproduction of charmonium states. This will be discussed in
sect.~\ref{VDM}. 

In the semiclassical Glauber approximation, we take into account these color
fluctuations in an effective way as described in ref.~\cite{farrar}. We
assume that a charmonium state $X$ is produced at $z$ as small $c\bar c$
configuration, then it evolves --
during the formation time $t_f$ resp. while it passes the formation length
$l_c$ -- to its full size. Please note that there is up to now no
theoretical or experimental proof for the assumption that charmonium states
are produced in point-like configurations as predicted in PQCD, a way to
test this experimentally was suggested recently in ref.~\cite{ger3}.
Therefore, if the formation length of the charmonium states, $l_f$, becomes
larger than the average internucleon distance ($l_f>r_{NN}\approx 1.8$ fm),
one has to take into account the evolution of the cross sections with the
distance from the production point $z'-z$~\cite{farrar}.
\begin{eqnarray}
\sigma(z,z')_X & = & \sigma(z)+{z'-z \over l_f}(\sigma_X-\sigma(z))
\;{\rm for}\;z'-z<t_f\; \cr
\sigma(z,z')_X & = &\sigma_X \hspace{4.1cm} {\rm otherwise}.
\label{expand}
\end{eqnarray}

The formation length of the $J/\psi$ is given by the energy denominator
$l_f\approx \frac{2p}{m^2_{\Psi'}- m^2_{J/\psi}}$, where $p$ is the momentum
of the $J/\psi$ in the rest frame of the target. With $p=30$ GeV, the
momentum of a $J/\psi$ produced at midrapidity at SPS energies
($E_{lab}=200$ AGeV), this yields $l_f\approx 3$ fm, i.e., a proper 
formation time of $\tau_f=0.3$ fm. 

As formation time of the $\psi'$ in its rest system we use the radius
given by nonrelativistic charmonium models, e.g. see the
refs.~\cite{quigg,eich}. This radius is $r=0.45$ fm for the $\psi$'.  Please
note that this radius differs from the variable $r$ in charmonium models,
which rather denotes the diameter of positronium like states.  Within the
formation time the cross section increases linearly with the distance from
the production point~\cite{farrar,ger}.

We chose a larger value for the cross section of the $\psi'$ N interaction
than for the $J/\psi$ because the radius of $\psi'$ is a factor two larger
and its radius is larger than the formation time of the $J/\psi$. A larger
value of $t_f$ for the $\psi'$ is supported also by the extraction of the
formation time of the $J/\psi$~\cite{kharzeev2}. 

Recently another attempt to describe the space-time evolution within the
formation time was published in ref.~\cite{volpe}. The authors develop a
quantum mechanical model to describe the expansion of the small wave
packages. Their initial condition is motivated by the NRQCD approach from
ref.~\cite{bodwin}. The charmonium states are described as superposition of
six charmonium states (4 $S$-waves and 2 $P$-waves). Diagonal and
nondiagonal transitions due to collisions with nucleons are taken into
account. In contrast to the GVDM in the model of ref.~\cite{volpe} partons 
interact with the nuclear target within the formation time. Because this 
model also leads to a stronger suppression at smaller Feynman x, the 
$c\bar c$ pairs within the formation time have a smaller absorption cross 
section than the fully formed charmonium states. This is, because the 
formation time decrease with the decrease of the Lorentz factor of the 
$c\bar c$ pairs relative to the nuclear target.

However, we restrict ourselves to the simpler model of
refs.~\cite{farrar,ger} because of simplicity and better control over the
impact. For a quantum mechanical description like in ref.~\cite{volpe} a
complete set of states is needed. The contribution of higher $S$ and $P$
states might be small, but the contribution of the continuum, given by
open charm states, is unclear (remember that the $\psi'$ is close to the
$D\bar D $ threshold, and a mixing with $D\bar D$ may be relevant for the
properties of large size configurations in the $\psi'$, 
cf.~\cite{Kogut:gr}).

The NRQCD approach for the production of
charmonium states is only valid if a large scale exists in the process,
e.g.\ a high transverse momentum. For $p_t$ integrated data, where small
transverse momenta dominate, this description doesn't work. This is
because~\cite{bodwin} colored states are produced that become color neutral
by radiation of gluons. Without a large scale in the process only soft
gluons can be emitted, which cannot transport quantum numbers (see   
e.g.~\cite{dok}). Another problem in ref.~\cite{volpe} is that a quantum
mechanical model is unable to describe the emission of gluons. Therefore in
this model the effect of the gluon emission on the $c\bar c$-distribution is
put in the initial distribution at the production point, though the emission
of soft gluons takes a relatively long time. 

In our ansatz these deviations from quantum mechanics predicted by quantum
field theories are taken into account in an effective way. In the expansion
of $c\bar c$ pairs suggested in ref.~\cite{farrar,ger} the cross sections 
of wave packages produced as small size objects increase with time, because 
the area of the color distribution increases. It is here not important if 
this increase is due to the motion of the partons or due to radiation of 
gluons. This scenario can be described like the diffusion of a $Q\bar Q$ pair 
in statistical mechanics, because the total cross section of such a pair
doesn't change due to gluon radiation if one sums over all channels,
i.e., over the number of emitted gluons~\cite{ddt}.

Ref.~\cite{benhar} claimed that the FNAL data on the $J/\psi$
photoproduction contradict to the expansion of $c\bar c$ pair discussed in
ref.~\cite{farrar}. But ref.~\cite{farrar} discussed the space-time
evolution of $c\bar c$ in the inclusive photoproduction of charmonium
states ($\gamma+A\to \Psi+ X$) for the energies where the coherence length
is $l_c\sim 2E_{\gamma}/M_{\psi}^2\ll R_A$. In the energy range of FNAL
data the coherence length is comparable or exceeds nuclear radius $R_A$. 
At high energies the $J/\psi$ is not produced locally as it is the case at
the SLAC energies considered in~\cite{farrar} and hence the analysis
of~\cite{farrar} is not literally applicable - one needs to take into
account both the formation of the $c\bar c$ before the nucleus and the
increase of the cross section of $c\bar c$-N interaction with energy. 
Besides ref.~\cite{benhar} discussed the quasielastic photoproduction of
charmonium states ($\gamma+A\to \Psi+p+(A-1)$). The selection of one
specific channel means that there is no summation over all radiated gluons
as for the inclusive process. Therefore the model of ref.~\cite{farrar}
should be modified to account for the suppression of gluon radiation.  the
expansion of the $c\bar c$ pair in the kinematics of FNAL photoproduction
data is described better by the generalized gluon distribution, i.e., by
the well understood QCD evolution ref.~\cite{Brodsky94,Collins96}. In this
paper we discuss the total charmonium-nucleon cross sections at moderately
large energies. Therefore the expansion model of ref.~\cite{farrar} should
be applicable.

\subsection{Vector Dominance Model\label{VDM}}

Somewhat different values for the cross section of the $J/\psi$ and
$\psi'$-nucleon interaction arise in charmonium models when the cross
section is assumed to be determined by the radius of the color
distribution within a hadron (tab.~\ref{meanb})  and also in the analysis
of photoproduction data within the Generalized Vector Dominance Model
(GVDM)~\cite{ger2}. The $\chi$-nucleon interaction can not be investigated
with this model, because it has different quantum numbers than the photon.
Therefore, parametrizations like in sect.~\ref{appendix} are needed to
make an educated guess for the $\chi$-nucleon cross section. 

The VDM takes into account only the direct diffractive production of the
$J/\psi$ and the $\psi'$, while the GVDM accounts also for the non-diagonal
transitions ($\psi'+N\rightarrow J/\psi+N$ and $J/\psi+N\rightarrow
\psi'+N$). The later are needed, because in photoproduction the particles
are produced as point like configurations and develop then to their average
size. In a hadronic model like the GVDM this is taken into account in form
of the interference due to the nondiagonal matrix elements. 

\begin{figure}
\centerline{\hbox{\epsfig{figure=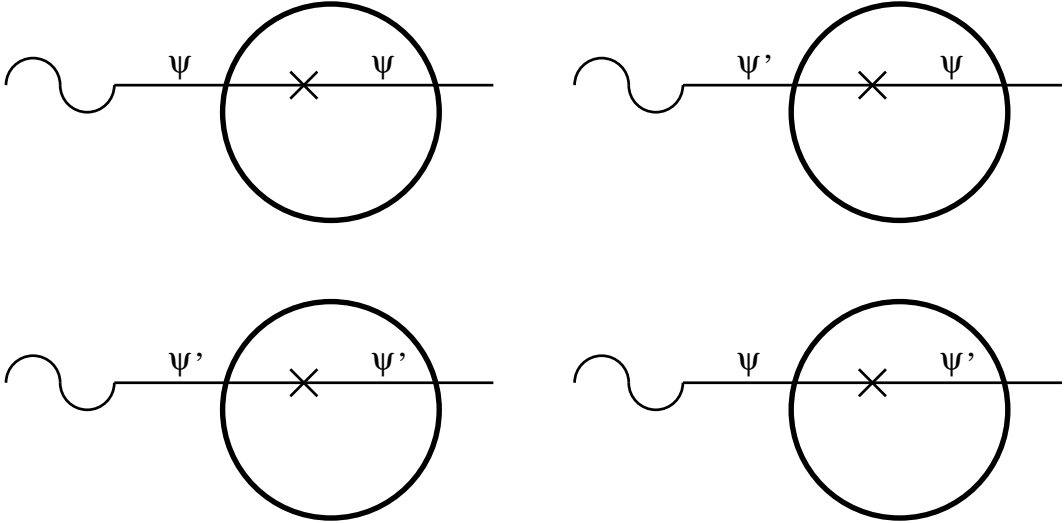,height=7cm}}}
\caption{Schematic illustration of the right hand side of eq.~(\ref{GVDM}).
}
\label{GVDMPIC}
\end{figure}

In the GVDM the photoproduction amplitudes $f_{\gamma \psi}$ and $f_{\gamma 
\psi'}$ for the $J/\psi$ and the $\psi'$
are given by~\cite{fs91}
\begin{eqnarray}
f_{\gamma \psi}&=&{e\over f_{\psi}}f_{\psi \psi}+{e\over f_{\psi'}}f_{\psi' 
\psi}\cr 
f_{\gamma \psi'}&=&{e\over f_{\psi}}f_{\psi \psi'}
+{e\over f_{\psi'}} f_{\psi' \psi'}
\label{GVDM}
\end{eqnarray}
$f_{\psi}$ and $f_{\psi'}$ are the $J/\psi-\gamma$ and the 
$\psi'-\gamma$ coupling and $f_{VV'}$ are the amplitudes for the processes 
$V+N\rightarrow V'+N$, where $V$ and $V'$ are the $J/\psi$ and the $\psi'$ 
respectively. In the VDM the non-diagonal amplitudes with $V\neq V'$ are 
neglected. The importance of the nondiagonal transitions is evident, because 
the left hand side of eq.~(\ref{GVDM}) is small. If it is neglected as a 
first approximation~\cite{fs91}, then $f_{\psi' \psi}=-{f_{\psi}\over 
f_{\psi'}} f_{\psi \psi}\approx 1.7\cdot f_{\psi \psi}$. And due to the
CPT-theorem $f_{\psi' \psi}=f_{\psi \psi'}$. The right hand side of 
eq.~(\ref{GVDM}) is illustrated in fig.~\ref{GVDMPIC}.

To get a real description of this expansion, it would be necessary to have
a complete set of hadron states, i.e., infinitely many bound S-wave
charmonium states plus the continuous spectrum (only S-waves have the same
quantum numbers as the photon). For illustrating the physics, we restrict
ourselves to the contribution of two states only. 

In ref.~\cite{ger2} the GVDM yields $8\div 10$ mb for the $\psi'$-nucleon
interaction cross section at SPS-energies. The sign of the photoproduction
amplitudes $f_{\gamma\psi}$ of the $J/\psi$ and $f_{\gamma \psi'}$ of the
$\psi'$is positive relative to $f_{\psi},\, f_{\psi'}$ within the
conventions of nonrelativistic charmonium models. The sign of the wave
functions in the origin are usually chosen to be positive.  It is easy to
see that a change of this convention would not change eq.~(\ref{GVDM}).
The $J/\psi$-nucleon interaction cross section at SPS-energies
approximately 3.5 - 4 mb is used as input into the analysis of
ref.~\cite{ger2}. The accuracy of such GVDM in predicting the $\psi'$N
cross sections is not clear. The above calculation demonstrates that
implementing color transparency leads to significantly larger cross
sections of the $\psi'$ N interaction. 

\subsection{Charmonium-Nucleon Cross Section within the Charmonium models
\label{appendix}}

One possibility to evaluate the predominantly nonperturbative QCD
contribution is to use an interpolation formula for the dependence of the
cross section on the transverse size $b$ of a quark-gluon configuration
within the constituent quark model,
\begin{equation}
\sigma_{abs}=c\cdot b^2
\label{constituent}
\end{equation}
where $b$ is the distance between the two constituent quarks, transverse to 
the collision direction. Such a form for the interpolation formula is also 
supported by pQCD, where the relation
\begin{equation}
\sigma(b)=\frac{\pi^2}{3}b^2\alpha_s(Q^2)\cdot xG(x,Q^2)\quad,
\label{sighard}
\end{equation}
$Q^2=9/b^2$, was found~\cite{sigma} for small $b^2$ only. Beyond small 
$b^2$ eq.~(\ref{constituent}) has no justification -- it is merely an
educated guess.

The constant $c$ can be adjusted within the constituent quark model:  
$b^2_{\pi}={8\over 3}r^2_{\pi}$ and $r^2_{\pi}$ is the square of the pion 
radius. This radius is known from  the Vector Meson Dominance Model to be 
$r_{\pi}={\sqrt{6}\over m_\rho}\approx 0.65$~fm~\cite{feyn}, 
$m_\rho\approx 770$~MeV is the mass of the $\rho$-meson. This result is 
confirmed by measurements of the electromagnetic form factor~\cite{na7} of 
the pion. Thus within this model
\begin{equation} 
c={\sigma_{\pi N}/ b^2_{\pi}}={25\,{\rm mb}/ 1.06\,{\rm fm}^2}=23.5 
{{\rm mb}/ {\rm fm}^2}
\end{equation}
We use in the following $\sigma_{\pi N}=25$ mb. This is in good agreement
with the data~\cite{pdg} for both energies discussed here
$E_{lab}(CERN)=450$~GeV and $E_{lab}(Hera\, B)=920$~GeV, because the energy
dependence of this cross section is not so strong in this kinematical
region. $b_{\pi}^2$ is increasing with energy as the cross section.  Within
soliton type models for the pion, the relation between $b$ and the radius of
the pion will be different. Therefore eq.~(\ref{constituent}) is model 
depended.

The $X-N$ cross sections is calculated via: 
\be
\sigma=\int \sigma(b)\cdot |\Psi (r)|^2 {\rm d}^3r \quad .
\label{integral}
\ee
where $\Psi (r)$ is the charmonium wave function. In our calculations
we use the wave functions from two non-relativistic charmonium models:
1.\ A Cornell confining potential, 
\begin{equation}
V_{Cornell}(r)=-\frac{0.52}{r}+\frac{r}{(2.34\,{\rm GeV}^{-1})^2}\,,
\end{equation}
see~\cite{eich} and refs. therein, and 2.\ a logarithmic potential
\begin{equation}
V_{log}(r)=-0.6635\mbox{ GeV}+(0.733\mbox{ GeV})\log(r\cdot 1\mbox{ GeV})
\end{equation}
from ref.~\cite{quigg}. We chose two different charmonium models to exhibit
the theoretical uncertainties of this approach. The resulting cross section
are given in table~\ref{meanb}. 

This also resolves the puzzle why in photon-nucleus collisions a smaller
($3\div 4$ mb) cross section for the $J/\psi$ was found than in
proton-nucleus collisions ($6\div 7.5$ mb): In pA collision 40\% of the genuine
$J/\psi$'s are from decays of $\chi$ mesons, which are strongly suppressed
in $\gamma$A collisions, because the $\chi$ has angular momentum $L=1$,
while $J/\psi$ and $\psi'$ has the same quantum numbers as the photon. 

\begin{table}[t]
\begin{center}
\caption
{\label{meanb}The average square of the transverse distances of the charmonium
states and the total quarkonium-nucleon cross sections $\sigma$ for two 
different charmonium models.
For the $\chi$ two values arise, due to the spin dependent wave functions
($lm=10,11$). The $\sigma_{hard}$ values are calculated with the
eqs.~(\ref{sighard}) and (\ref{integral}). $b\leq 0.35$ is an upper limit
for the integral in eq.~(\ref{integral}). For the row $\sigma(hard, all\,
b)$ we used a parametrization like eq.~(\ref{sighard}) above $b=0.9$~fm,
because in this region Bjorken-$x$ becomes larger than 1 and the gluon 
distribution of eq.~(\ref{sighard}) cannot be defined in this region.} 
\begin{tabular}{|c|cccc|}
\hline
$c\overline{c}$/$b\overline{b}$-state & J/$\Psi$ & $\Psi'$ & $\chi_{c10}$ & 
$\chi_{c11}$ \\ 
\hline
$<b^2>$ (fm${}^2$) & 0.094 & 0.385 & 0.147 & 0.293 \\
\hline
$\sigma(Cornell)$ (mb) & 2.2 & 9.1 & 3.5 & 6.9 \\
\hline
$\sigma(log.)$ (mb) & 2.7 & 13 & 4.4 & 8.7 \\
\hline
$\sigma(hard,b<0.35\mbox{ fm})$& 0.42 & 0.13 & 0.41 & 0.28 \\
\hline
$\sigma(hard, all\, b)$& 2.2 & 8.2 & 3.7 & 7.4 \\
\hline
\end{tabular}  
\end{center}
\end{table}

\subsection{Hard cross section\label{hardxsect}}

We also calculate $\sigma(X-N)$ assuming that pQCD is
applicable. In this calculation we ignore the differences
between bare quarks of the QCD Lagrangean and the constituent quarks,
because
these are nonperturbative QCD effects.  The numerical calculations shown in
Tab.~\ref{meanb} yields at CERN energies $\sigma(X-N)$ values which are
significantly smaller than in~\cite{slac}. The $\sigma_{hard}$
values~\cite{foot} in Tab.~\ref{meanb} are also
calculated with eq.~(\ref{integral}), but here we integrate only over the
region 0 fm $<b<0.35$ fm, because $\alpha_s$ increases with $b$. $b$ is
the {\it transverse distance} between the heavy quarks transverse to the
momentum of the heavy quarkonium.  For the row $\sigma(hard, all\, b)$ we
used a parametrization like eq.~(\ref{sighard}) above $b=0.9$~fm, because in
this region Bjorken-$x$ becomes larger than 1: the gluon distribution of
eq.~(\ref{sighard}) cannot be defined in this region! For the calculation of
the hard cross section, the $\sigma(b)$ in eq.~(\ref{integral}) is given by
eq.~(\ref{sighard}). We employ the CTEQ5L parameterizations of the gluon
distribution functions in the proton~\cite{cteq}.

The Bjorken x needed for the calculation of $\sigma_{hard}(X-N)$ is
calculated by $x={Q^2 / (2m_N\nu)}$, where $Q^2={9/ b^2}$, $m_N$ is the
nucleon mass and $\nu$ is the energy of the heavy Quarkonium state $X$ in
the rest frame of the nucleon. The calculation in Tab.~\ref{meanb} is done
for a state $X$ produced at midrapidity for SPS energies, but in the target
fragmentation region for RHIC and LHC. One can see that the hard
contributions to the cross sections are just a correction at SPS energies,
but at RHIC energies both contributions become compatible, while at LHC the
hard contributions dominate (we neglect here that the DGLAP
(Dokshitzer-Gribov-Lipatov-Altarelli-Parisi) equation might be
violated~\cite{felix}). 

The extrapolation 
of eq.~(\ref{sighard}) yields similar values as a parametrization like
eq.~(\ref{constituent}). This is because the factor $\alpha_s\cdot
xG(x,Q^2)$ depends only weakly on $x$ and $Q^2$, even in the region where
$Q^2\ll 1$ GeV${}^2$. This is no prove that parametrizations like
eq.~(\ref{constituent}) can be applied to charmonium states, but at least it
shows that there is no discontinuity between the pQCD result and the
parametrizations for the soft regime.

\section{Experimental results}

\subsection{Comparisson with CERN Data}

\begin{figure}
\centerline{\hbox{\epsfig{figure=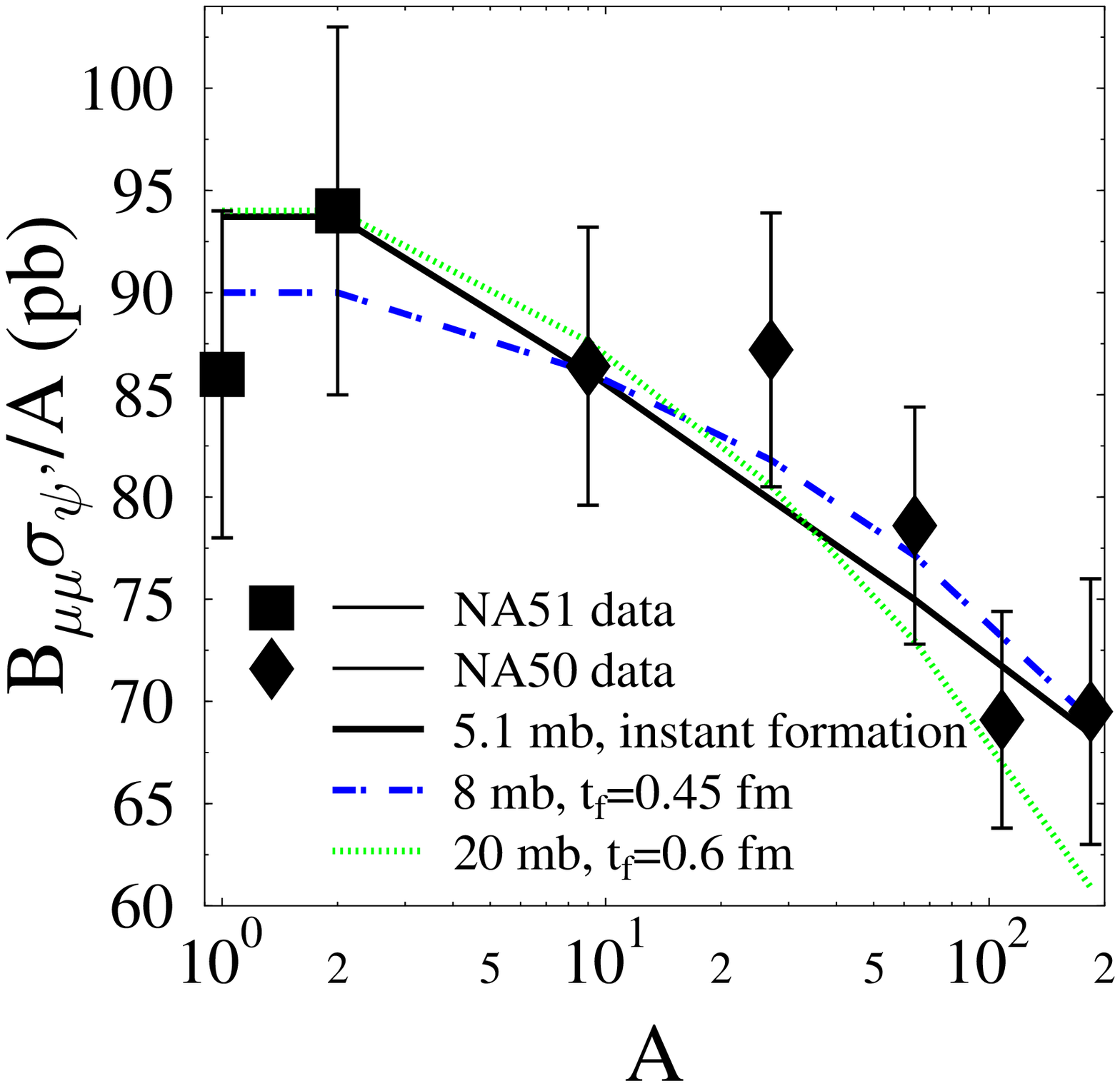,height=9cm}\epsfig{figure=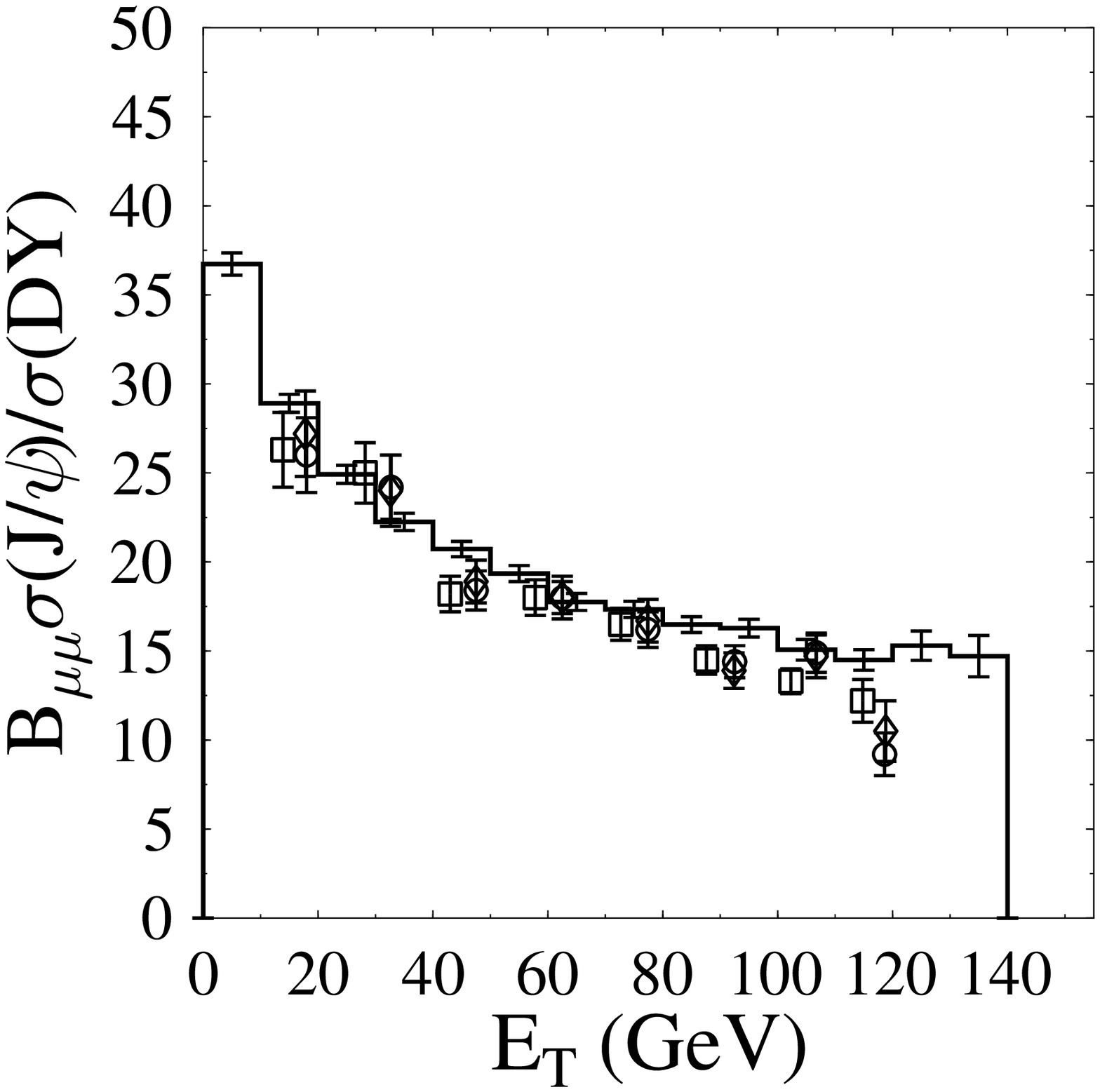,height=9cm}}}
\caption{
On the (lhs) are plotted the $B_{\mu\mu}\sigma_{\psi'}/A$ values extracted
from calculations with different absorption cross sections and different
formation times, the NA51 data for pp and pD, and the NA50 data for pBe,
pCu, pAg, and pW vs. the mass $A$ of the target. And the (rhs) shows
$B_{\mu\mu}\sigma(J/\psi)/\sigma(DY)$ with the absorption cross sections
from ref.~\protect\cite{ger} and the NA50 data for PbPb collisions vs. the
transverse energy $E_t$.
}
\label{na50s}
\end{figure}

In the lefthand side of fig.~\ref{na50s} we show a comparison between 
calculations with different
cross sections and different expansion times and the NA50 data~\cite{na50}
for pA collisions and the NA51 data~\cite{na51} for pp and pD collisions for
the cross section of $\psi'$ N interaction vs. the mass of the target. The
y-axis shows $B_{\mu\mu}\sigma_{\psi'}/A$ where $B_{\mu\mu}$ is the
branching ratio for the decay of the $\psi'$ into dimuon pairs. and
$\sigma_{\psi'}$ is the production cross section. The "5.1 mb, instant
formation"  curve in fig.~\ref{na50s} (lhs) is the fit of the NA50 
collaboration
to their data. Instant formation means that they assumed that the $\psi'$ is
produced with the full cross section and not as a point like particle as in
the description of this paper. (Note the the NA50 collaboration fitted
$B_{\mu\mu}\sigma_{\psi'}/\sigma_{DY}$, where $DY$ means Drell-Yan,
therefore we multiplied there fit with the Drell-Yan cross section in pp
collisions measured by NA51). 

The "8 mb, $t_f=0.45$~fm" curve is the eye-ball fit of the model described
in this paper. For the comparison with the data we need the production
cross section of the $\psi'$ in pp collisions as input. We used here the
average of the pp and pD data of the NA51 collaboration. The value of 8 mb
agrees well with the model parametrizations discussed in the
sections~\ref{model} and ~\ref{model2}. However, we compare also with the
calculation with the parameters of ref.~\cite{ger}, i.e.,
$\sigma(\psi'N)=20$~mb and $t_f=0.6$~fm. For this comparison we used the
production cross section of the $\psi'$ in pD collisions divided by two as
input. This is also close to the value of the NA50 fit. One can see in the
lefthandside of fig.~\ref{na50s} that the calculation with these
parameters is also in good agreement with the data. A value for
$\sigma(\psi'N)$ of the size of 20 mb is favored by the nucleus-nucleus
data as shown in ref.~\cite{spieles}. 

In the righthandside of fig.~\ref{na50s} is plotted
$B_{\mu\mu}\sigma(J/\psi)/\sigma(DY)$ the NA50 data~\cite{na502} for PbPb
collisions vs. the transverse energy $E_t$, a measure for the centrality
of the collision. The result of the calculation within the UrQMD
(Ultrarelativistic Quantum Molecular Dynamics)  model~\cite{bass} is also
shown in form of the histogram in fig.~\ref{na50s} (rhs). In the
calculation (for more details see ref.~\cite{spieles}) for the charmonium
nucleon absorption cross sections the values from ref.~\cite{ger} were
used. In this calculation the charmonium states can interact also with
secondary produced particles.  Based on the additive quark model the cross
section for charmonium baryon interactions was assumed to be the same as
the charmonium nucleon cross sections and for charmonium meson
interactions two thirds of the charmonium nucleon cross sections. 

The calculation for the $J/\psi$ agrees well with the data. The calculation
for the $\psi'$ underestimates the data. However it is not understood, if
this is due to the high value of $\sigma(\psi'N)=20$~mb, or if nondiagonal
transitions like in sect.~\ref{VDM} should be taken into account in AA
collisions, too.

The value of 8 mb for the $\psi'$ nucleon cross section from the fit to the
pA is smaller than the theoretical estimate 20 mb of ref.~\cite{ger}. This
is because in ref.~\cite{ger} a formation time of 0.6 fm was chosen for the
$\psi'$, while we used here 0.45 fm, the radius of the $\psi'$ given by the
charmonium models. The fact that the formation time is not known very well
is another uncertainty.  Further uncertainty comes from using the diffusion
model of expansion at distances comparable to the scale of the soft
interaction. Within the error bars the $\psi'$-nucleon cross section
extracted from these pA data and the prediction of the GVDM, discussed in
sect.~\ref{VDM}, are qualitatively similar. However, further data are needed
to learn more about this cross section.

\subsection{Predictions for HERA B\label{HERA B}}

\begin{figure}
\centerline{\hbox{\epsfig{figure=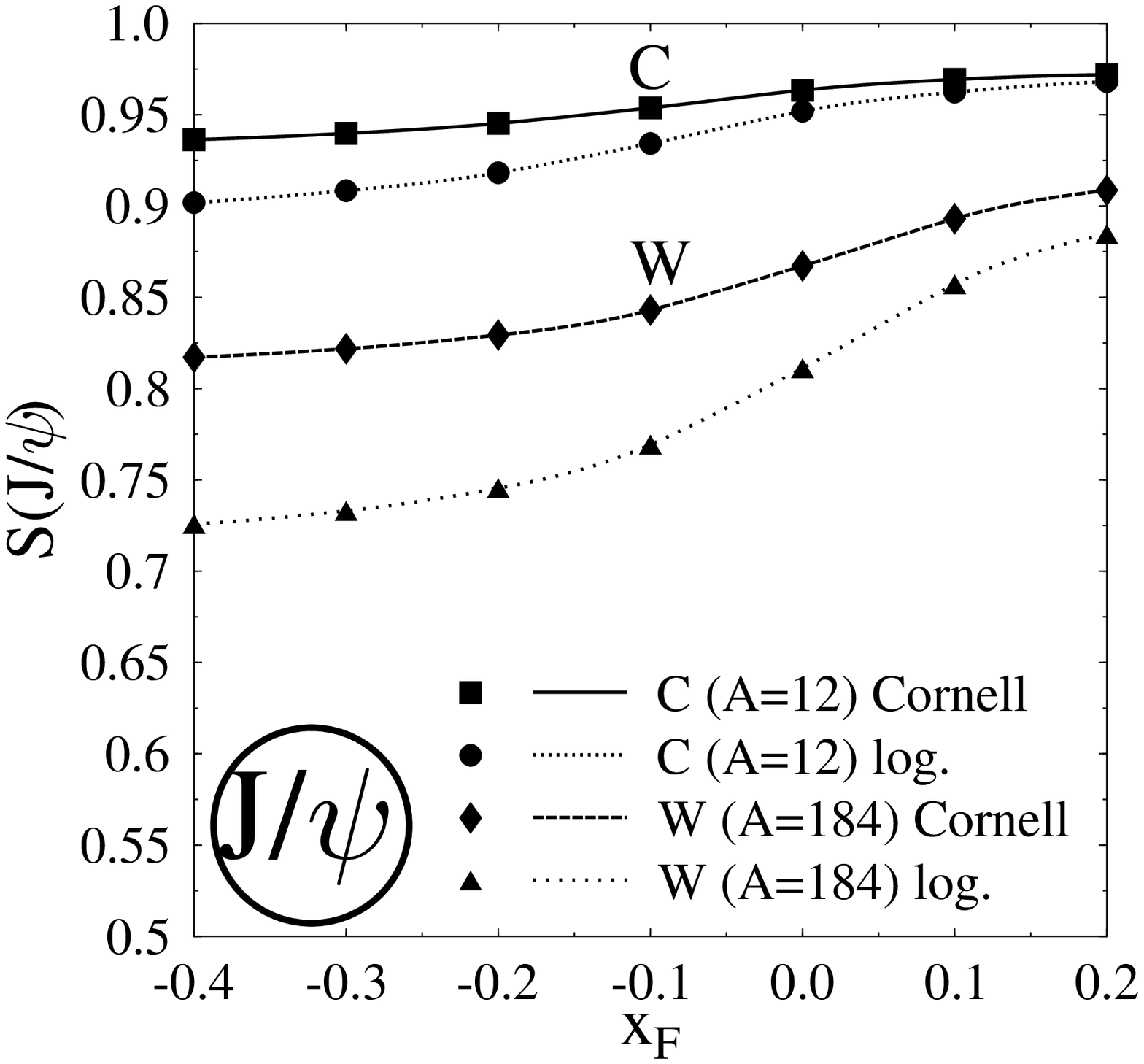,height=9cm}\epsfig{figure=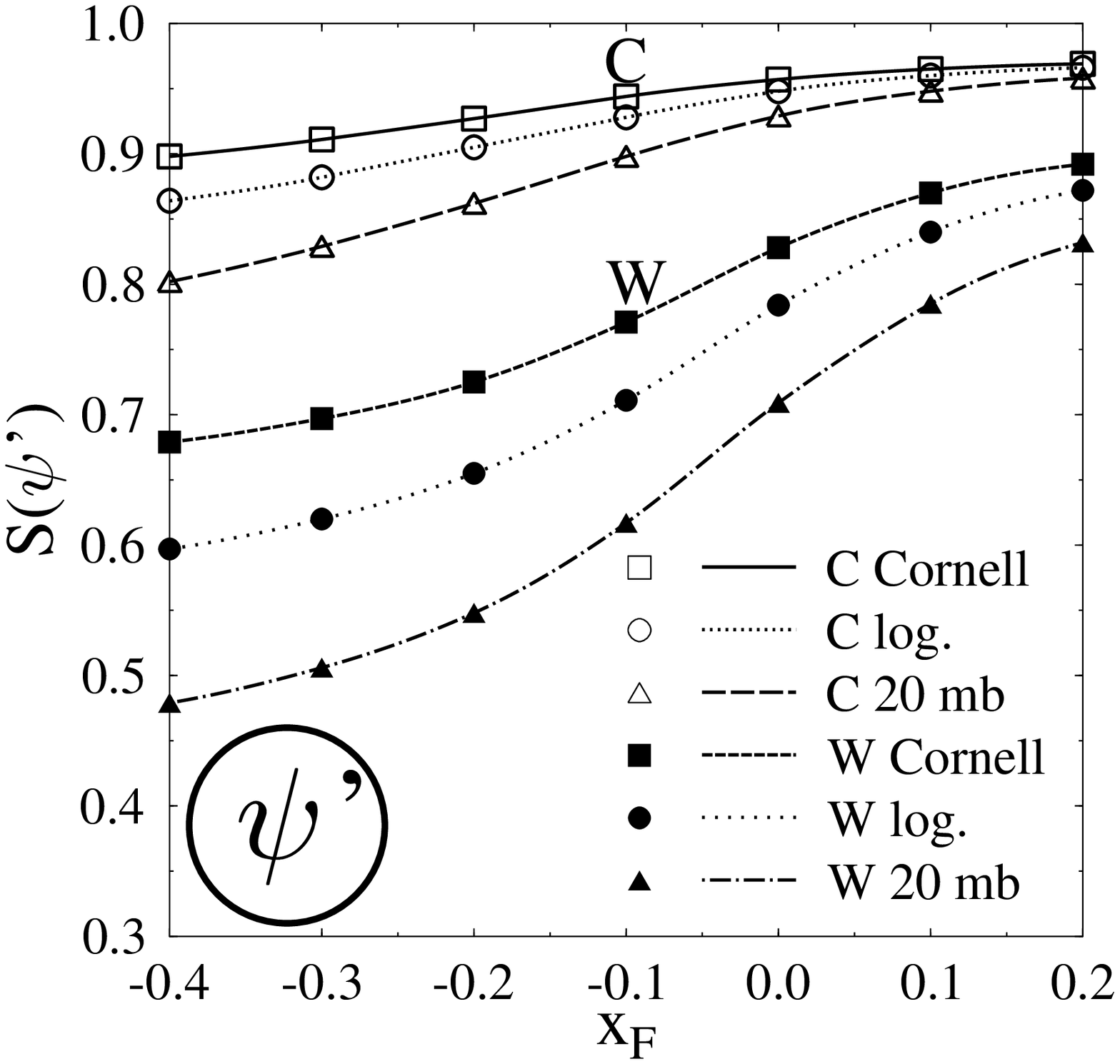,height=9cm}}}
\caption{$S(J/\psi)$ (lhs) and $S(\psi')$ (rhs) are shown vs. $x_F$ for
Carbon and Tungsten at HERA B 
energies ($E_{lab}=920$ GeV).}
\label{papsi}
\end{figure}

Fig.~\ref{papsi} shows our predictions of $S(J/\psi)$ lefthand side (lhs) 
and $S(\psi')$ (rhs) for HERA B vs.
Feynman x for pA collisions at $E_{lab}=920$ GeV. Two different nuclear
targets, Carbon ($A=12$) and Tungsten with ($A=184$), are shown for the two
different sets of cross sections resulting from the different charmonium
models. 

We used here the cross sections of the constituent quark model shown in
tab.~\ref{meanb} in sect.~\ref{appendix}, because this model is the only one 
which predicts also $\chi$-nucleon cross sections. 
We calculated the suppression factor $S$ for the $\psi'$ additionally with a
cross section of 20 mb and an formation time of 0.6 fm to estimate
theoretical uncertainties.  These are the values discussed in
ref.~\cite{ger}.

Note that we plotted in fig.~\ref{papsi} the result of our model up to a
Feynman x of 0.2 to show the behavior of the model.  However, it is known
from E866~\cite{leitch} and NA3~\cite{na3} data that there are additional
suppression mechanisms at positive Feynman x, e.g.\ parton energy loss and
nuclear shadowing.

\section{Conclusions}

The new data of the NA50-collaboration~\cite{na50} and the data of the
E866-collaboration~\cite{leitch} prove that the $\psi'$-nucleon cross
section is much larger than the $J/\psi$-nucleon cross section.  This is in
agreement with the photoproduction data for these charmonium states as
discussed in the framework of the GVDM in section~\ref{VDM}. This confirms
the QCD prediction that the strength of hadron-hadron
interactions depends on the volume occupied by color.

Within the assumption that charmonium states are produced as point like
white states, we demonstrated that the data~\cite{na50}
can be fitted with a $\psi'$-nucleon cross section of
$\sigma(\psi'N)\approx 8$~mb. We discuss different models for this cross 
section and show that they prefer a $\psi'$-nucleon cross section of
$\sigma(\psi'N,\mbox{ model})=9\div 13$~mb. However, due to the large
experimental errors we conclude that the data and the QCD-motivated models
agree, further data with higher accuracy and covering a larger rapidity range
are needed. 

This cross section will be measured soon in proton-nucleus collisions
at HERA B at an energy of $E_{lab}=920$~GeV. The advantage of this
experiment is that it covers a larger range of Feynman-$x_F$, especially
in the negative $x_F$ region. In this region effects due to the formation
time of the hadron will be less important. These data will
give new information about the relations between the size of mesons and the
strength of hadron-hadron interactions as well as the possibility to test
the existing models for this relation.

\vspace*{1cm}
{\bf Acknowledgement:}\\
L.G.\ acknowledges support by the Minerva Foundation. H.S.\
acknowledges support by BMBF, DFG and the GSI. L.F. acknowledges
support by GIF, M.S. is acknowledges support by DOE.

%\end{references}


\begin{thebibliography}{99}

\bibitem{na50} R. Shahoyan (NA50) Contributed to 37th Rencontres de 
Moriond on QCD and Hadronic Interactions, Les Arcs, France, 16-23 Mar 2002 
%arXiv:hep-ex/0207014\\
B.~Alessandro {\it et al.}  [NA50 Collaboration],
%``Charmonia And Drell-Yan Production In Proton Nucleus Collisions At The
%Cern Sps,''
Phys.\ Lett.\ B {\bf 553}, 167 (2003).
%%CITATION = PHLTA,B553,167;%%

\bibitem{ger}L.~Gerland, L.~Frankfurt, M.~Strikman, H.~St\"ocker and W.~Greiner,
%``J/psi production, chi polarization and color fluctuations,''
Phys.\ Rev.\ Lett.\  {\bf 81} (1998) 762\\
%[arXiv:nucl-th/9803034].
%%CITATION = NUCL-TH 9803034;%%
L.~Gerland, L.~Frankfurt, M.~Strikman, H.~St\"ocker and W.~Greiner
Nucl. Phys. {\bf A663} (2000) 1019 
%[arXiv:nucl-th/9908052].
%%CITATION = NUCL-TH 9908052;%%

\bibitem{leitch}
M.~J.~Leitch {\it et al.}  [FNAL E866/NuSea collaboration],
%``Measurement of J/psi and psi' suppression in p A collisions at
%800-GeV/c,''
Phys.\ Rev.\ Lett.\  {\bf 84} (2000) 3256
%[arXiv:nucl-ex/9909007].
%%CITATION = NUCL-EX 9909007;%%

\bibitem{sigma} L. Frankfurt, A. Radyushkin, M. Strikman
Phys. Rev. {\bf D55} (1997) 98 \\
B. Bl\"attel {\it et al.}, Phys. Rev. Lett. {\bf 71} (1993) 896\\
L. Frankfurt, G. Miller, M. Strikman, Phys. Lett. {\bf B304} (1993) 1

\bibitem{slac} R. L. Anderson {\it et al.},
Phys.\ Rev.\ Lett.\  {\bf 38} (1977) 263

\bibitem{eich} E. Eichten {\it et al.}, Phys. Rev. {\bf D21} (1980), 203

\bibitem{quigg} C. Quigg and J. L. Rosner, Phys. Lett {\bf B71} (1977) 153

\bibitem{spieles}C.~Spieles, 
R.~Vogt, L.~Gerland, S.~A.~Bass, M.~Bleicher, H.~St\"ocker and 
W.~Greiner, Phys.\ Rev.\ C {\bf 60} (1999) 054901
%[arXiv:hep-ph/9902337].
%%CITATION = HEP-PH 9902337;%%

\bibitem{yennie} D.R.\ Yennie, Hadronic Interactions of Electrons and Photons,
Academic (1971) 321

\bibitem{devries} C.W. deJager, H. deVries, and C. deVries
{\it Atomic Data and Nuclear Data Tables} {\bf 14} (1974) 485

\bibitem{kharzeev} D. Kharzeev, C. Lourenco, M. Nardi, H. Satz:
Zeit. Phys. {\bf C74} (1997) 307

\bibitem{FS85} L. Frankfurt and M. Strikman Nucl\ Phys.\ {\bf B250} (1985) 
147

\bibitem{fs91} L.~Frankfurt, M.~Strikman, Progress in Particle and Nuclear 
Physics 27 (1991) 136

\bibitem{farrar} G. Farrar, L. Frankfurt, M. Strikman, H. Li
Phys.\ Rev.\ Lett. {\bf 64} (1990) 2996

\bibitem{ger3}
L.~Frankfurt, L.~Gerland, M.~Strikman and M.~Zhalov,
%``The psi-prime-prime to psi-prime ratio as unambiguous signature for hard 
%physics in nuclear reactions and in decays of beauty hadrons,''
Phys. Lett. {\bf B563} (2003) 68
%%CITATION = HEP-PH 0302009;%%

\bibitem{kharzeev2}
D.~Kharzeev and R.~L.~Thews,
Phys.\ Rev.\ C {\bf 60} (1999) 041901

\bibitem{volpe}
D.~Koudela and C.~Volpe,
%``Charmonium production in relativistic proton nucleus collisions: What  
%will we learn from the negative x(F) region?,''
arXiv:hep-ph/0301186.
%%CITATION = HEP-PH 0301186;%%

\bibitem{bodwin}
G.~T.~Bodwin, E.~Braaten and G.~P.~Lepage,
%``Rigorous QCD analysis of inclusive annihilation and production of heavy
%quarkonium,''
Phys.\ Rev.\ D {\bf 51}, 1125 (1995)
[Erratum-ibid.\ D {\bf 55}, 5853 (1997)]
%[arXiv:hep-ph/9407339].
%%CITATION = HEP-PH 9407339;%%

\bibitem{Kogut:gr}
J.~B.~Kogut and L.~Susskind,
%``Electron - Positron Annihilation At And Above The Charm Threshold,''
Phys.\ Rev.\ Lett.\  {\bf 34}, 767 (1975).

\bibitem{dok}
Yu.~L.~Dokshitzer, "Bremstrahlung", contribution to the ASI sommer school
2000, at Advanced Study Institute (St. Croix).

\bibitem{ddt}
Yu.~L.~Dokshitzer, D.~Diakonov and S.~I.~Troian,
%``Hard Processes In Quantum Chromodynamics,''
Phys.\ Rept.\  {\bf 58}, 269 (1980).
%%CITATION = PRPLC,58,269;%%

\bibitem{benhar}
O.~Benhar, B.~Z.~Kopeliovich, C.~Mariotti, N.~N.~Nicolaev and
B.~G.~Zakharov,
%``Why Photoproduction Of Charmonium On Nuclei Does Not Measure The
%Charmonium Nucleon Total Cross-Section,''
Phys.\ Rev.\ Lett.\  {\bf 69}, 1156 (1992).
%%CITATION = PRLTA,69,1156;%%

\bibitem{Brodsky94}
S.~J.~Brodsky, L.~Frankfurt, J.~F.~Gunion, A.~H.~Mueller and M.~Strikman,
%``Diffractive leptoproduction of vector mesons in QCD,''
Phys.\ Rev.\ D {\bf 50} (1994) 3134
%[arXiv:hep-ph/9402283].
%%CITATION = HEP-PH 9402283;%%
%\cite{Collins:1996fb}

\bibitem{Collins96}
J.~C.~Collins, L.~Frankfurt and M.~Strikman,
%``Factorization for hard exclusive electroproduction of mesons in QCD,''
Phys.\ Rev.\ D {\bf 56} (1997) 2982
%[arXiv:hep-ph/9611433].
%%CITATION = HEP-PH 9611433;%%

\bibitem{ger2}L.~Gerland, L.~Frankfurt, M.~Strikman, M.~Zhalov 
%``Cross section oscillations in the coherent charmonium photoproduction  
%off nuclei at moderate energies,''
arXiv:hep-ph/0301028 and
%%CITATION = HEP-PH 0301028;%%
%``Probing coherent charmonium photoproduction off light nuclei at medium 
%energies,''
arXiv:hep-ph/0301077.
%%CITATION = HEP-PH 0301077;%%

\bibitem{feyn} R.~Feynman 'Photon-Hadron Interaction', Addison-Wesley

\bibitem{na7} 
S.~Amendolia {\it et al.}, Nucl. Phys. {\bf B277} (1986), 168

\bibitem{pdg} K. Hagiwara {\it et al.}, Phys. Rev. {\bf D66} (2002) 010001 

\bibitem{foot}
The calculated $\sigma_{hard}$ reflects the effective cross
section for the interaction of a $Q\bar Q$ pair with nucleons of the
residual nucleus, which will transform into the corresponding $Q\bar Q$
state only after it has traversed the nucleus.

\bibitem{cteq}
H.~L.~Lai {\it et al.}  [CTEQ Collaboration],
Eur.\ Phys.\ J.\ C {\bf 12}, 375 (2000).
%[arXiv:hep-ph/9903282].
%%CITATION = HEP-PH 9903282;%%

\bibitem{felix} FELIX Collaboration (E. Lippmaa {\it et al.}):
CERN-LHCC-97-45, Aug 1997, 197pp

\bibitem{na51}
M.~C.~Abreu {\it et al.}  [NA51 Collaboration],
%``J/Psi, Psi' And Drell-Yan Production In P P And P D Interactions At
%450-Gev/C,''
Phys.\ Lett.\ B {\bf 438}, 35 (1998).
%%CITATION = PHLTA,B438,35;%%

\bibitem{na502}
P.~Cortese {\it et al.} [NA50 Collaboration],
Nucl. Phys. {\bf A715}, 243 (2003) 

\bibitem{bass}
S.~A.~Bass {\it et al.},
%``Microscopic models for ultrarelativistic heavy ion collisions,''
Prog.\ Part.\ Nucl.\ Phys.\  {\bf 41}, 225 (1998)
%[arXiv:nucl-th/9803035].
%%CITATION = NUCL-TH 9803035;%%

\bibitem{na3}
J.~Badier {\it et al.}  [NA3 Collaboration],
%``Experimental J / Psi Hadronic Production From 150-Gev/C To 280-Gev/C,''
Z.\ Phys.\ C {\bf 20}, 101 (1983).
%%CITATION = ZEPYA,C20,101;%%

\end{thebibliography}
\end{document}